\documentclass[acmlarge]{acmart}
\AtBeginDocument{%
  }
\usepackage{multirow}
\usepackage{graphicx}
\usepackage{subcaption}
\usepackage{booktabs}
\graphicspath{ {./images/} }
\setcopyright{cc}
\setcctype{by}
\acmJournal{IMWUT}
\acmYear{2025} \acmVolume{9} \acmNumber{1} \acmArticle{217}
\acmMonth{3} \acmPrice{}\acmDOI{10.1145/3712284}
\acmISBN{978-1-4503-XXXX-X/2018/06}

\begin{document}

\title[Watch Out! E-scooter Coming Through]{Watch Out! E-scooter Coming Through!: Multimodal Sensing of Mixed Traffic Use and Conflicts Through Riders’ Ego-centric Views} 


\author{Hiruni Nuwanthika Kegalle}
\affiliation{%
  \institution{RMIT University}
  \city{Melbourne}
  \country{Australia}}
\email{hiruni.kegalle@student.rmit.edu.au}
\orcid{0009-0002-9758-1368}

\author{Danula Hettiachchi}
\affiliation{%
  \institution{RMIT University}
  \city{Melbourne}
  \country{Australia}}
\email{danula.hettiachchi@rmit.edu.au}
\orcid{0000-0003-3875-5727}

\author{Jeffrey Chan}
\affiliation{%
  \institution{RMIT University}
  \city{Melbourne}
  \country{Australia}}
\email{jeffrey.chan@rmit.edu.au}
\orcid{0000-0002-7865-072X}

\author{Mark Sanderson}
\affiliation{%
  \institution{RMIT University}
  \city{Melbourne}
  \country{Australia}}
\email{mark.sanderson@rmit.edu.au}
\orcid{0000-0003-0487-9609}

\author{Flora D. Salim}
\affiliation{%
  \institution{University of New South Wales}
  \city{New South Wales}
  \country{Australia}}
\email{flora.salim@unsw.edu.au}
\orcid{0000-0002-1237-1664}

\renewcommand{\shortauthors}{Kegalle et al.}

\begin{abstract}
E-scooters are becoming a popular means of urban transportation. However, this increased popularity brings challenges, such as road accidents and conflicts when sharing space with traditional transport modes. An in-depth understanding of e-scooter rider behaviour is crucial for ensuring rider safety, guiding infrastructure planning, and enforcing traffic rules. In this paper, we investigated the riding behaviours of e-scooter users through a naturalistic study. We recruited 23 participants, equipped with a bike computer, eye-tracking glasses and cameras, who traversed a pre-determined route, enabling the collection of multi-modal data. We analysed and compared gaze movements, continuous speed, and video feeds across three different transport infrastructure types: a pedestrian-shared path, a cycle lane and a roadway. Our findings reveal that e-scooter riders face unique challenges, including difficulty keeping up with faster-moving cyclists and motor vehicles due to the capped speed limit on shared e-scooters, issues in safely signalling turns due to the risks of losing control when using hand signals, and limited acceptance from other road users in mixed-use spaces. Additionally, we observed that the cycle lane has the highest average speed, the least frequency of speed change points, and the least head movements, supporting the suitability of dedicated cycle lanes -- separated from motor vehicles and pedestrians -- for e-scooters. These findings are facilitated through multimodal sensing and analysing the e-scooter riders' ego-centric view, which show the efficacy of our method in discovering the behavioural dynamics of the riders in the wild. Our study highlights the critical need to align infrastructure with user behaviour to improve safety and emphasises the importance of targeted safety measures and regulations, especially when e-scooter riders share spaces with pedestrians or motor vehicles. The dataset and analysis code are available at \url{https://github.com/HiruniNuwanthika/Electric-Scooter-Riders-Multi-Modal-Data-Analysis.git}. 
\end{abstract}





\maketitle

\section{INTRODUCTION}
Micro-mobility is emerging as a significant mode of urban transport due to its various benefits. Micro-mobility promotes environmental sustainability, mitigates traffic congestion, and fosters healthier lifestyles \cite{USDepartmentofTransportation, MOURATIDIS2021102716}. With the proliferation of micro-mobility solutions, there is a critical need for an in-depth understanding of rider behaviour as a way to address road safety issues. Previous research shows the importance of examining rider behaviour to ensure rider safety \cite{BENDAK2021269}, improve infrastructure \cite{WESTERHUIS2017146}, adopt practices \cite{9304835}, and enforce traffic rules \cite{behaviorallyComplexEscooter}. 
E-scooters have become a prominent area of research for several reasons: (1) They introduce a novel experience for both riders and other road users who interact with them, presenting unique challenges and dynamics within public spaces \cite{10.1145/3313831.3376499}. (2) They are permitted on multiple transport infrastructure, that are allocated for traditional transport modes \cite{10.1145/3313831.3376499,10.1145/3491101.3519831}. (3) Although e-scooters are often considered similar to bicycles, they differ in terms of usage patterns and social acceptance \cite{10.1145/3491101.3519831, almannaa2021comparative}. It is evidenced by recent discussions at UBICOMP \cite{10.1145/3570345, 10.1145/3610904} and CHI \cite{10.1145/3491101.3519831, 10.1145/3313831.3376499,10.1145/3334480.3375169}.

Traditional methods such as interviews \cite{doi:10.1080/17450101.2021.1967097}, survey \cite{LAA2020102874}, and observations \cite{anke2023micro} have been commonly employed for the analysis of rider behaviour. However, these approaches are subject to limitations, including potential recall bias and subjectivity in responses \cite{burt2023scooter}. By integrating real-time speed, gaze behaviour data and ego-centric video analysis with traditional methods, we aim to obtain a more accurate and comprehensive understanding of rider behaviour. Although previous research has successfully used multi-modal data collection methods, such as video feeds, accelerometer and eye-tracking data \cite{10.1145/3568444.3568451, 6595621, PETZOLDT2017477, 7795977} to analyse cyclists' behaviour, these studies primarily focused on bicycles. 

Cities adopt different infrastructure policies for e-scooters, allowing them to operate on various types of transport infrastructure \cite{bai2020dockless, foissaud2022free}. Given the novelty of e-scooters, there is still uncertainty surrounding optimal infrastructure adaptation and rider integration. This variation in infrastructure influences rider behaviour, infrastructure adaptability, and interactions with other road users. However, studies on e-scooter rider behaviour across various types of infrastructure remain limited. To address this gap, we conduct a comparative analysis of rider behaviour across three different infrastructure types.

Inspired by existing naturalistic studies on cycling behaviour, we conducted a study to investigate the variations in e-scooter rider behaviour across different transport infrastructures. Our approach combined quantitative and qualitative analyses to examine various aspects of e-scooter rider behaviour. Specifically, we compared continuous speed, gaze movements, head movements, and fixated Areas of Interest (AOIs) of e-scooter riders when navigating three distinct types of transport infrastructure: pedestrian-shared paths, cycle lanes, and roadways. We used Tobii Pro 3 Glasses, Garmin Edge 130 Plus bike computer, Insta 360 camera, and GoPro HERO 10 Camera for data collection. The main contributions of our work include:
\begin{itemize}
    \item We collected a multi-modal dataset through a naturalistic study converging various mobility infrastructures. The features characterising rider behaviour (e.g., speed, eye movement, head movements, gestures) were identified from the literature and appropriate devices to capture each of these were selected. To ensure reproducibility and support future micro-mobility modeling, we have made the dataset (excluding collected videos due to privacy concerns) and analysis code publicly available\footnote{\url{https://github.com/HiruniNuwanthika/Electric-Scooter-Riders-Multi-Modal-Data-Analysis.git}}.
    \item We applied a combination of multi-modal data analysis methods, established in the literature, to the novel context of e-scooter use. This approach enabled us to derive valuable insights into e-scooter rider behaviour, interactions, conflicts, and responses of other road users.
\end{itemize}

The rest of the paper is structured as follows. Section 2 distinguishes e-scooter riding behaviour from cycling and reviews the existing literature. In Section 3, we present the naturalistic study design followed by the data analysis process. The results of our study are presented in Section 4. Section 5 offers a discussion of the findings. In Section 6, limitations and future work are discussed. Finally, we conclude the paper in Section 7. 
\section{RELATED WORK}

    \subsection{Distinct Characteristics of E-Scooter Rider Behaviour}

Although extensive research has explored cyclist behaviours and safety perceptions, the emergence of e-scooters as a novel mode of transport requires a specific investigative focus. Distinct from bicycles, e-scooters exhibit unique usage characteristics that demand specific attention in research. The \textit{demographic profile} of e-scooter users differs from that of cyclists. According to \cite{WeRideAustraliaReport}, the majority of e-scooter users are predominantly within the age groups of 18-34 and 35-44, whereas cycling participation is considerably higher among older age groups. Specifically, only 13.46\% of individuals aged between 45 and 54 engage in e-scooter usage, compared to 37\% in the same age group who participate in cycling. This trend is supported by additional research \cite{su142114303, NIKIFORIADIS2021102790}, which indicates that younger individuals are more inclined to use e-scooters, thereby the rider behaviour of cyclists and e-scooter riders can be different.

\textit{Accident statistics} reveal that incidents involving e-scooters surpass those associated with bicycles. A comparative study by \citet{james2023comparison} on injuries related to e-scooters, bicycles, and motorbikes, found that injuries from e-scooter use surged by 2.8 times over a four-year period, while bicycle-related injuries increased by only 1.2 times. This marked disparity highlights the need for further research into understanding user behaviour specific to e-scooters. Furthermore, \citet{su11205591} highlights that individuals feel less comfortable around e-scooters compared to bicycles. This might exhibit varied pedestrian reactions that influence differences in rider behaviours.

The \textit{manoeuvring patterns} of e-scooters, which are powered electrically, differ significantly from those of traditional bicycles \cite {10.1145/3313831.3376499}. For instance, research by \citet{visualAttenstionTobii} demonstrated that e-scooters are more likely to weave through pedestrian traffic compared to bicycles. \citet{dozza2022data} suggest that e-scooters offer greater manoeuvrability and comfort than bicycles, while they necessitate longer braking distances.

Unlike bicycles, \textit{regulations} governing e-scooters exhibit significant variability across localities \cite{10.1145/3313831.3376499, NIKIFORIADIS2021102790}. For example, Italian law restricts e-scooter speeds to 20 km/h, permits their use on both pedestrian and cycle lanes, and sets a minimum user age of 14 years \cite{su142114303}.  In contrast, Austrian regulations require a minimum rider age of 12 years, enforce a maximum speed limit of 25 km/h, and stipulate that e-scooters be used on bike paths or roadways where bike paths are not available \cite{CountryOverview}. This diversity is not only international but also within a country, reflecting the challenges of integrating this novel transportation mode into existing systems. Due to their recent introduction, the precise impact of e-scooters remains unclear. Therefore, some cities conduct e-scooter trials before integrating e-scooters in their transport systems \cite{VICtrial, AusTrials}.

In summary, in contrast to bicycles, which are well-established components of transportation planning, e-scooters represent unique challenges that emphasise the importance of e-scooter-focused rider behaviour analysis.

    \subsection{Traditional Methods for Capturing Rider Behaviour}
There are multiple methodologies employed to study e-scooter rider behaviour and interactions. \textit{(1) Interview} is one of the most common approaches that provides detailed insights into e-scooter usage \cite{10.1007/978-3-030-85613-7_26,10.1145/3313831.3376499, doi:10.1080/17450101.2021.1967097}. However, the presence of an interviewer might influence responses, and participants may not fully disclose their behaviours due to biases like social desirability or recall issues \cite{ burt2023scooter}. 
\textit{(2) Observations} \cite{TUNCER2020102702, 10.1145/3313831.3376499,10.1145/3544548.3581049, 10.1145/3544548.3581045, anke2023micro} on individual rider allow researchers to directly assess rider behaviour, which might be affected by observer bias. Further, the data analysis can be complex and time-consuming. \textit{(3) Survey} \cite{LAA2020102874, GIOLDASIS2021106427, anke2023micro, WEISS2024100047} provide data from a broad sample of riders but are prone to personal biases and often lack the context of traffic conditions that might influence riding behaviour \cite{10.1145/3025453.3025911}. \textit{(4) Media report analysis} \cite{GOSSLING2020102230} offers a broader public opinion on issues and trends, yet it typically focuses more on accidents rather than normal riding behaviors, which might skew understanding of everyday e-scooter use \cite{YANG2020105608, WHITE2023182}.

These traditional methods have strengths and inherent limitations in different situations. Overall, they can be influenced by personal biases, inaccuracies in self-reporting, lack of context details and can be time-consuming. In contrast, the use of sensors and cameras to capture events as they occur can provide an unbiased and detailed record of behaviours that might be unnoticeable to the human eye.

    \subsection{Using Multi-modal Data to Analyse Rider Behaviour}

The usage of different data modalities such as video \cite{hong2022evaluation, 10.1145/3313831.3376499,KOVACSOVA2018270, KAYA2021106380, WHITE2023182}, gaze movements \cite{visualAttenstionTobii, 10.1145/3544548.3581049, MANTUANO2017408,10.1145/3204493.3214307, KAYA2021106380, VONSTULPNAGEL2020222}, physiological measurements \cite{COBB2021172, PhysiologicalCyclist, distefano2020physiological},
and vehicle operation data (e.g., speed, acceleration, braking, steering angle) \cite{MORGENSTERN2020104740, doi:10.1080/15389588.2019.1643015} to understand driver or rider behaviour has been an established area of research (see Table \ref{tab:lit_gap}).

\begin{table}[h]
    \centering
    \captionsetup{justification=centering}
    \caption{Related works for rider behaviour analysis with different data modalities.}
    \label{tab:lit_gap}
    \small
    \begin{tabular}{lccccccccc}
        \toprule
         & \multicolumn{4}{c}{Road user} & & \multicolumn{3}{c}{Device} & \\
         \cmidrule(lr){2-5} \cmidrule(lr){7-9}
         Paper & Pedestrian & Cyclist & E-scooter & Driver & & Acceler & Eye & Video   & Infrastructure \\
         & & & rider & & & -ometer & -tracker & camera & comparison\\
        \midrule
        \citet{10.1145/3313831.3376499} & & & \checkmark & & & & & \checkmark & -  \\ \hline
        \citet{hong2022evaluation} & & & \checkmark & & & & & \checkmark  & -  \\ \hline
         \citet{visualAttenstionTobii} & \checkmark  & \checkmark  & \checkmark & & & \checkmark & \checkmark & & -  \\ \hline
         \citet{10.1145/3544548.3581049} & & \checkmark & & & & & \checkmark & & -  \\ \hline
         \citet{10.1145/3568444.3568451}  & & \checkmark & & & & & \checkmark & \checkmark & Yes  \\ \hline
        \citet{MANTUANO2017408} & & \checkmark & & & & & \checkmark & & Yes \\ \hline
        \citet{6595621} & & \checkmark & & & & \checkmark & & \checkmark & - \\ \hline
        \citet{PETZOLDT2017477} & & \checkmark & & & & \checkmark & & \checkmark & - \\ \hline
        \citet{7795977} & & \checkmark & & & & \checkmark & & \checkmark & - \\ \hline
        \citet{10.1145/3204493.3214307} & \checkmark & \checkmark & & & & & \checkmark & & - \\ \hline
       \citet{KAYA2021106380} & & & & \checkmark & & & \checkmark & \checkmark & -  \\ \hline
        Ours & & & \checkmark & & & \checkmark & \checkmark & \checkmark & Yes \\
        \bottomrule
    \end{tabular}
\end{table}

E-scooters being a novel transport modality, researchers have conducted qualitative video-based studies to understand the rider behaviour. \citet{10.1145/3313831.3376499} explored the riding practices of e-scooter riders, employing a video ethnographic study. They collected video recordings of 3 e-scooter riders in urban spaces, using a chest-mounted camera worn by a researcher. Their findings revealed information on negotiations with other users of public spaces, and the novel adaptation of riding behaviour in various circumstances, such as traffic lights and road crossings. \citet{hong2022evaluation} used video recordings collected from fixed cameras located at different sites (e.g., side walks, cross walks, intersections, pedestrian areas) to analyse e-scooter traffic. They reported instances of hazardous riding behaviours and the ways in which e-scooter riders modify their riding strategies in response to varying traffic conditions.

Given that eye movements indicate visual attention, eye-tracking based studies have been used to understand road users' behaviour \cite{10.1145/3607822.3614532}. \citet{visualAttenstionTobii} compared the visual attention and speed of 12 participants while they were functioning as pedestrians, cyclists, and e-scooter riders on a shared road. The results showed the distribution of visual fixations across AOIs varied between the different types of users. Specifically, cyclists and e-scooter riders were more likely to focus their attention on the road ahead as compared to pedestrians. \citet{10.1145/3544548.3581049} collected gaze movements and GPS data from 12 commuter cyclists. Their analysis showed how cyclists' gaze patterns are distributed across different parts of vehicles, traffic lights and road markings in various traffic contexts. To compare the movement patterns of cyclists and pedestrians, \citet{10.1145/3204493.3214307} used gaze movement data and accelerometer data collected from eye-tracker glasses. They found that both cyclists and pedestrians have common gaze sequences and different patterns of shoulder checks. \citet{KAYA2021106380} explored the effect of drivers’ cycling experience on their visual scanning behaviour using a vehicle instrumented with two cameras, and eye movements of the driver. Their results showed that drivers with cycling experience have a lower probability of visual scanning failures compared to other drivers.

Only a few naturalistic studies have experimented with the differences in rider behaviours across various types of transport infrastructure \cite{MANTUANO2017408, 10.1145/3568444.3568451}. \citet{MANTUANO2017408} examined the gaze behaviour of cyclists on cycle tracks, both exclusive and shared with pedestrians, involving 16 participants. They analysed the proportion of fixations and attention dispersion using eye-tracker glasses. In contrast to their approach, our study collects a more diverse range of data and considers additional traffic conditions, thereby enabling a broader set of evaluation metrics. \citet{10.1145/3568444.3568451} explored the behaviour of cyclists on self-balancing bicycles across different types of transport infrastructure, including parks, main roads, and junctions, through video footage and eye movement analysis. Different to our research, which primarily examines the natural riding behaviours of e-scooter users, their study focuses on the multitasking behaviours of participants while riding cycles.

In summary, different from previous efforts, our work compares the e-scooter rider behaviour at different transport infrastructure, in contrast to other studies that focus on rider behaviour in specific traffic situations such as left turns, traffic light-protected intersections, and cross walks. Furthermore, we employ additional devices to collect more data, allowing us to obtain multiple measurements for improved comparisons. Our research specifically targets e-scooter riders, as opposed to cyclists, pedestrians or drivers, thereby shedding light on this novel mode of transport users.

\section{STUDY}

We conducted a naturalistic study to explore the e-scooter rider behaviour on different infrastructure types. Each participant was riding on a pre-determined path, wearing a selected set of equipment, including eye-tracking glasses and a helmet-mounted camera. The riding behaviour of the participant was recorded using a chest-mounted camera worn by another following rider. This study design was chosen so that participants could exhibit their natural riding behaviour and genuine interaction patterns.

\begin{figure}[ht]
  \centering
  \begin{subfigure}[b]{0.48\textwidth}
    \includegraphics[width=\textwidth]{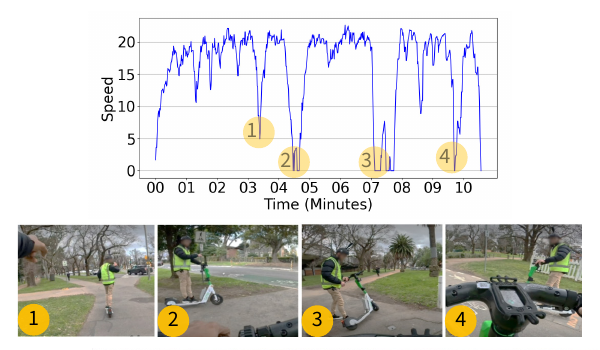}
    \caption{Round 1.}
    \label{fig:round_1_final}
  \end{subfigure}
  \hfill  
  \begin{subfigure}[b]{0.48\textwidth}
    \includegraphics[width=\textwidth]{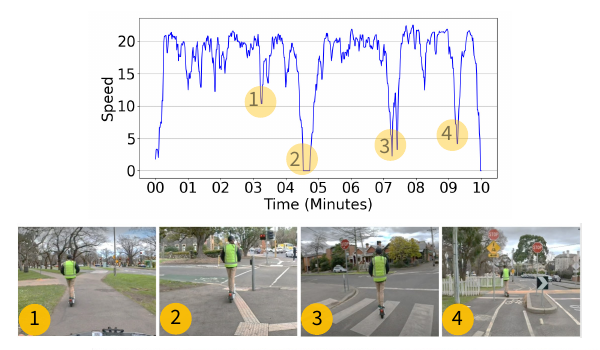}
    \caption{Round 2.}
    \label{fig:round_2_final}
  \end{subfigure}
    \caption{Different behaviours of P7 at turning points.}
\end{figure}

To closely mimic real-world riding conditions and minimise the impact of unfamiliarity bias on the results, we asked the participants to complete two rounds on the pre-determined route. This approach was informed by survey findings \cite{Lime_Micromobility} indicating that e-scooters are predominantly used for commuting, typically along familiar routes. 
The two example images, Figures \ref{fig:round_1_final} and \ref{fig:round_2_final}, demonstrate the efficacy of our study design. These figures compare speed variability and associated incidents for P7, across two rounds. Notably, in the second round, there is notable stabilisation in speed and smoother turning behaviour, suggesting increased familiarity with the route. 

        \subsection{Devices}
We used Tobii Pro Glasses 3 eye-tracker\footnote{\url{https://go.tobii.com/tobii-pro-glasses-3-user-manual}} to record gaze movements. It stores the video footage in first-person view, captured at a 25 frames-per-second (FPS) rate. The glasses were lightweight ($312$ $g$), and the portable control unit, which collects data, made it easy to use while riding. Earlier versions of Tobii Glasses have been used in in-the-wild studies to collect eye movements \cite{visualAttenstionTobii, 10.1145/3204493.3214307}.

Participants wore a helmet-mounted Insta360 X3 camera\footnote{\url{https://www.insta360.com/product/insta360-x3}}. It recorded the rider's surroundings with a resolution of $5952 \times 2976$ pixels at a 30 FPS rate. We selected the Insta360 camera due to its ability to capture a complete 360-degree view surrounding the participant, thereby offering us extensive and detailed visual data.

We equipped the participant's e-scooter with a Garmin Edge 130 Plus bike computer\footnote{\url{https://www.garmin.com/en-AU/p/698436}}, which collected GPS and speed data. The lightweight and compact design ensures no added inconvenience to the rider. The bike computer has an extended battery life, and the data can be exported through a web portal.

The following rider (i.e., researcher) captured the behaviour of the participant using a chest-mounted GoPro HERO 10 camera\footnote{\url{https://gopro.com/en/au/shop/cameras/hero10-black/CHDHX-101-master.html}}. It recorded the video footage at 60 FPS with a resolution of $5312 \times 2988$ pixels. This has commonly been used in in-the-wild experiments \cite{10.1145/3313831.3376499}. 

\begin{figure}[h]
  \centering
  \includegraphics[width=0.65\textwidth]{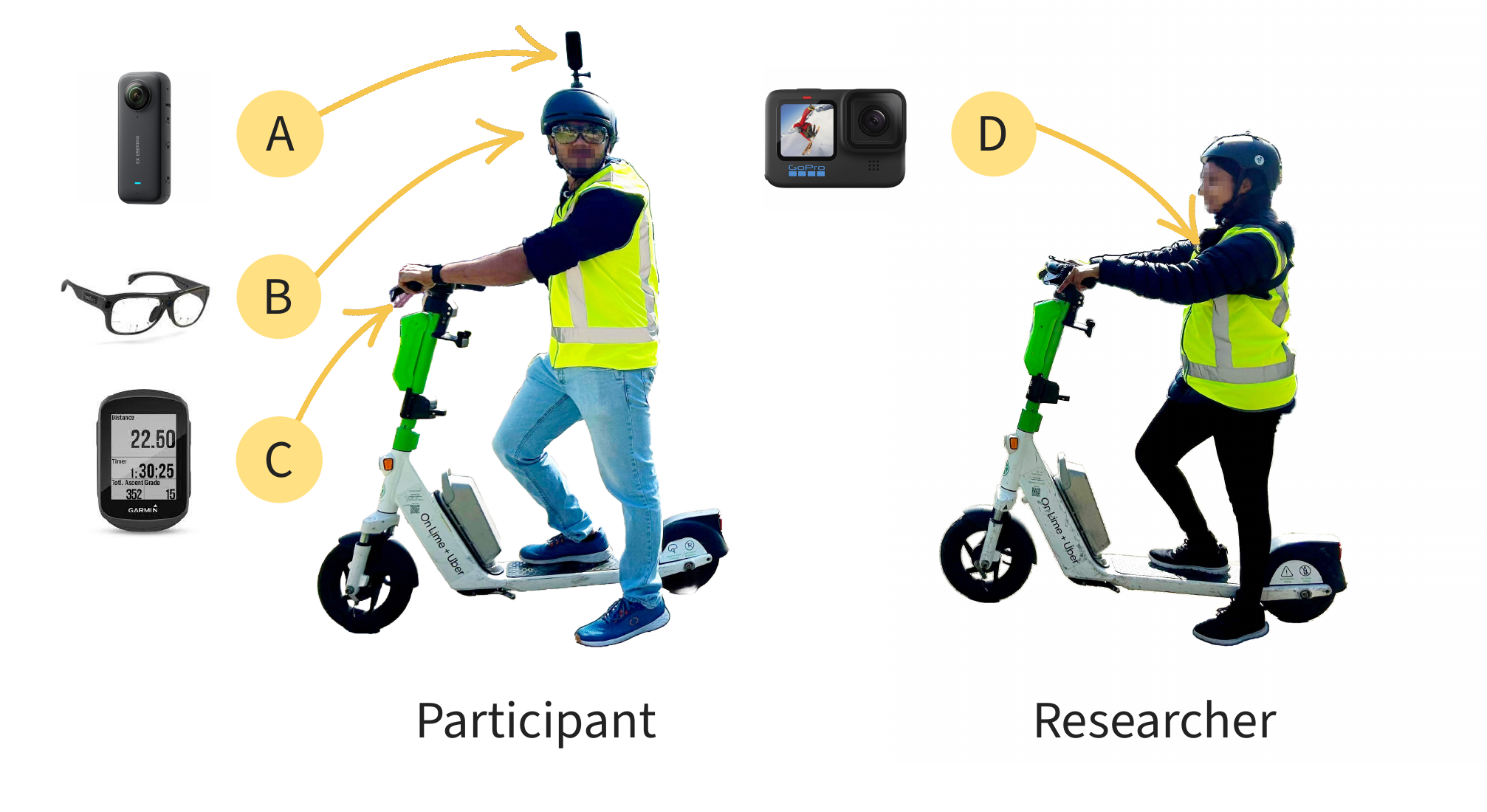}
  \caption{Participant equipped with (A) Insta 360 camera, (B) Tobii Pro 3 Glasses, (C) Garmin bike computer and a following rider equipped with (D) GoPro Camera.}
  \label{fig:deviceSetup}
\end{figure}

We calibrated all the devices to local timestamp, facilitating the data integration process. Both the participant and the following rider used regular shared e-scooters operated by one of the provider companies involved in the city’s trial. The study setup, including the devices, the participant and the following rider, is illustrated in Figure~\ref{fig:deviceSetup}.

        \subsection{Task}
Each participant was asked to ride the e-scooter and given directions of the route using a map. We asked them to ride naturally while adhering to the local road rules. While the participant rides, a member of the research team follows them in another e-scooter. The following rider was instructed to alert the rider only if she diverts from the pre-determined route. 

The task, as designed, uses a naturalistic observation approach \cite{angrosino2016naturalistic} since we capture the interactions in their natural environment without any intervention. This approach was chosen primarily to understand the behaviour of both riders and other road users at encounters. 

        \subsection{Study Procedure}
Prior to data collection, we gave the participants a brief introduction to the study objectives, the equipment involved, and the data to be captured. We also provided a document outlining local traffic regulations about e-scooter use. Then, we verified the participant's age using identification documents, and obtained their verbal confirmation on prior e-scooter riding experience. Next, we assisted participants in wearing the helmet-mounted camera, and eye-tracking glasses. We calibrated the eye tracker following the standard procedure, and gave them time to get familiar with the glasses. Finally, a designated safety officer, part of the research team, inspected the device setup. 

Following the inspection, we attached the bike computer to the participant's e-scooter. To comply with insurance, study participants used their own shared e-scooter mobile application\footnote{\url{https://www.li.me/en-au/the-app}} to initiate the ride (At the end of the study, we reimbursed each participant for the trip cost incurred during the study). We asked them to unlock the e-scooter using the e-scooter App. Next, we gave them a brief acclimatisation period, during which they rode the e-scooter in a relatively safe test zone. Lastly, we obtained verbal confirmation from participants that they were comfortable and confident riding the e-scooter with the study equipment.

After completing the preparation phase, we started recording data on all four devices, including the GoPro camera worn by the following rider. Once the first round was finished, we stopped the recordings. We began the second round after re-calibrating the eye tracker and starting the record. At the end of each participant's ride, we invited them for a short follow-up interview to share their feelings about the study and discuss any noteworthy interactions they encountered.

        \subsection{Participants}
All the e-scooter riders recruited for the study had corrected or uncorrected 6/6 vision. The participants comprised 13 men (18-34 years old = 10, 35-54 years old = 3) and 10 women (18-34 years old = 7, 35-54 years old = 3). Even though there is a gender and age imbalance, our study sample aligns with reports on typical e-scooter user demographics \cite{Contributor_2023}. Among all the participants, 9 were frequent e-scooter users (2-3 days a week), and 14 were using it rarely. Participants were recruited based on the responses to a flyer displayed on social media and compensated with an AUD 50 gift voucher for their effort and time. The study was approved by the RMIT University Human Research Ethics Committee\footnote{Ethics Approval Number: 26049.}, and all the riders were covered with two insurance covers provided by the university and the e-scooter service provider company.

        \subsection{Route and Interactions}
\begin{figure}[htb]
  \centering
  \includegraphics[width=0.3\textwidth]{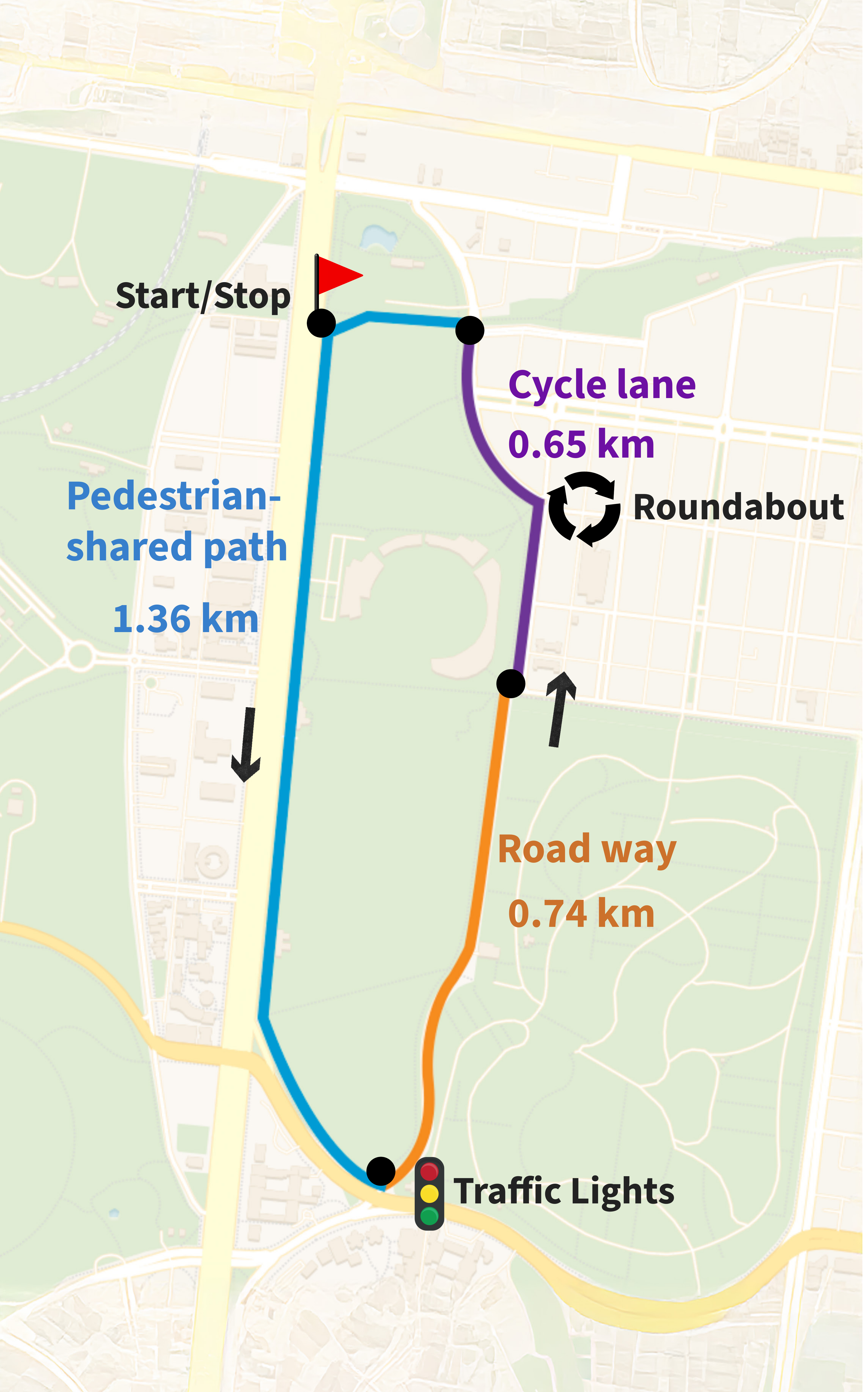}
  \caption{The route.}
  \label{fig:route}
\end{figure}

The route (See Figure~\ref{fig:route}) featured three distinct types of road infrastructure, chosen for several key reasons. E-scooters are a relatively new addition to transport networks, and cities are still determining the most suitable road environments for them. As discussed in the literature, different cities have adopted varying infrastructure policies for e-scooter use. In Melbourne, Australia, where this study was conducted, local regulations permit e-scooters to operate on pedestrian-cyclist shared paths, cycle lanes, and roadways with speed limits below 60 km/h when no dedicated cycle lane is available. To comply with these regulations and to explore the most relevant road environments, we selected these three infrastructure types. This selection also allowed us to examine the variability of rider behaviour, as different road environments influence factors such as speed, caution, and decision-making. This provided a comprehensive understanding of how e-scooters interact with various road conditions. Additionally, by selecting a variety of road types, the study explored infrastructure adaptability, shedding light on how e-scooter riders adjust to changing environments -- critical for cities integrating e-scooters into mixed-use spaces. Furthermore, this selection facilitated an analysis of the impact on other road users, such as pedestrians, cyclists, and motor vehicles. Understanding these interactions across different infrastructure types is essential for informing road-sharing strategies and conflict mitigation measures. For example, on pedestrian-cyclist shared paths, participants encountered both pedestrians and cyclists, while on roadways, they interacted with moving and parked vehicles. By selecting these three types of infrastructure, we maximized the diversity of rider behaviour and road user interactions.

The first route segment features a shared pedestrian-cycle path, flanked by trees on the left and a parallel roadway on the right, with an additional walking path visible beyond the trees. It includes a few cross streets where motor vehicles are allowed, which introduces additional traffic elements into the path. The second segment is a cycle lane, accompanied on the right by an opposing-direction cycle lane and on the left a separated vehicle lane with parking slots. The third section is a roadway (mixed vehicle-cycle lane), characterised by parked vehicles on the left and moving vehicles on the right. This section contains one roundabout, three stop signs, and a pedestrian crossing, adding complexity to the rider's decision-making process. The final segment completes the route by connecting to the starting point, and we have not considered that in the analysis. In the city where the experiment was conducted, e-scooter speed restriction of 20 km/h was imposed on all types of roads.

During this period, four participants encountered atypical traffic conditions due to a local event near the study area, which altered typical traffic dynamics along the route. On the pedestrian-shared path, the usual average of 34 pedestrians increased to 67, with one to three motor vehicles observed -- normally absent in this segment. In the cycle lane, pedestrian and cyclist numbers remained stable despite the event. The roadway, however, saw a notable rise in both stationary and moving motor vehicles, with counts nearly doubling from normal conditions. These variations highlight the inherent unpredictability of naturalistic field studies \cite{visualAttenstionTobii}.

        \subsection{Measures}

\begin{figure}[h]
  \centering
  \includegraphics[width=0.9\textwidth]{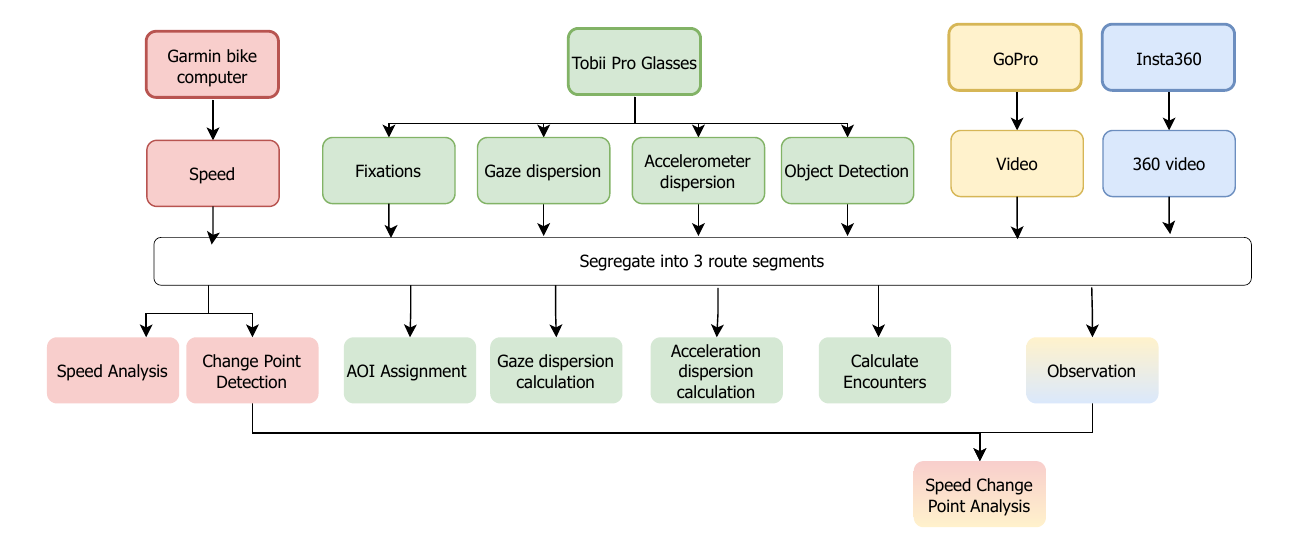}
  \caption{Sensor data analysis workflow.}
  \label{fig:workflow}
\end{figure}

In this study, we employed a mixed-methods approach, integrating both qualitative and quantitative assessment techniques to comprehensively analyse the data. The qualitative evaluation was conducted through the analysis of video feeds, providing insights into participant behaviours and environmental interactions. For the quantitative analysis, we compared various metrics across different route segments, including speed, fixation counts, and encounters. The analysis workflow is shown in Figure \ref{fig:workflow}.
            
            \subsubsection{Quantitative Evaluation}
Garmin Edge 130 Plus bike computer recorded the riding speed at one-second intervals. We extracted data using Garmin Connect\footnote{\url{https://connect.garmin.com/modern}}, after which the data were converted to include local timestamps. Then, the data were segregated into three route segments. We cleaned the dataset following the common procedure of removing stand-still situations (speed=0 km/h) \cite{SCHLEINITZ2017290}.
 
To identify speed change points, we partitioned the speed time series into multiple intervals that exhibited variations in speed, utilizing the pruned exact linear time (PELT) \cite{PELTalgorithm} change point detection method, which has been previously employed for similar analyses in the literature \cite{YAN201839}. Accelerations and decelerations during the ride were detected using the PELT algorithm. 

Data extraction from the eye-tracking glasses was conducted using Tobii Pro Lab software\footnote{\url{https://www.tobii.com/products/software/behavior-research-software/tobii-pro-lab}}. We annotated the data corresponding to each segment of the route and calculated the average fixation count. Consistent with the methodology outlined in \cite{10.1145/3568444.3568451}, we computed the average values for gaze standard deviation based on fixations in the x and y axes. Although gaze dispersion is typically measured in angles, in \cite{10.1145/3568444.3568451} it was calculated as the Euclidean distance using normalised pixel values; we adopted the same approach for our analysis.
Furthermore, following the guidelines of \citet{10.1145/3568444.3568451}, we computed the average values for the standard deviation of accelerometer data, reflecting head movements along the x, y, and z axes. Similarly to the method used for gaze dispersion, we calculated the Euclidean distance using normalised pixel values, adhering to the established protocol in \cite{10.1145/3568444.3568451}.

The gaze movements are used to detect the rider's glances, which indicates their attentional focus. We analysed eye movement data using the fixation-by-fixation method, as detailed in previous studies \cite{doi:10.1080/00140139.2014.990524, visualAttenstionTobii}. Tobii Pro Lab software, equipped with the Tobii Pro Velocity-Threshold Identification Gaze Filter, facilitated the identification and visualisation of fixation points. For each identified fixation, we extracted the surrounding region ($radius = 70 pixels$) using OpenCV. To analyse the contents within these regions, we employed a pre-trained Region-Based Convolutional Neural Network (R-CNN) \cite{NIPS2015_14bfa6bb}. However, the automated object detection yielded suboptimal accuracy levels, with mean pedestrian detection recall at 10.5\% and mean car detection recall at 15.1\%. To address this limitation, we subsequently conducted a manual annotation process, by assigning each fixation to its respective AOI. To mitigate researcher bias, the annotation process was conducted collaboratively, involving two researchers to ensure a balanced and unbiased assessment.

We used YOLO v8.0 model \cite{yoloCode} for the automatic analysis of video data. This enabled us to contextualize the study route and its traffic conditions. We extracted the number of pedestrian, cyclist and car encounters in each of the segments. The model accuracy of the object detection was confirmed with manual annotation of 7 randomly selected participant videos. The model accuracy for pedestrian, cyclist and motor vehicle detection is 74.4\%, 71.4\%, and 82.9\%, respectively.

                \subsubsection{Qualitative Evaluation}

To conduct qualitative observations of rider behaviour and the reactions of other road users in different route segments, we utilised video feeds from Insta 360 and GoPro cameras. The 360-degree videos were reviewed using Insta360 Studio software\footnote{\url{https://www.insta360.com/download/insta360-x3}}, which facilitates a multi-directional viewing capability.

We employed thematic analysis \cite{doi:10.1080/2159676X.2019.1628806, alma9922061070801341} to identify, analyse, and interpret patterns across the video recordings. This approach was selected to capture the complexity of interactions between e-scooter riders and other road users, focusing on recurring themes related to behaviour and communication. As demonstrated in previous studies, thematic analysis has been successfully applied to explore interaction patterns in similar contexts \cite{madigan2019understanding, mcilroy2021thinking}.

Two researchers collaboratively reviewed three video recordings and coded each interaction scenario. The coding process focused on rider behaviour, responses of other road users, and communication methods employed during these encounters. This initial round of coding generated a preliminary codebook, which was used as a guideline for analysing the remaining videos. As the analysis progressed, the codebook was updated to incorporate new observations and emerging insights.

\section{RESULTS}
    \subsection{Quantitative}
        \subsubsection{Speed}
Due to data unavailability (partially recorded), we could only analyse speed data from 19 participants. We annotated data according to the 3 segments of the route and calculated average speed (See Table \ref{tab:avg_speed}). According to Shapiro-Wilk tests, the average speed follows a normal distribution. Results of the repeated measures ANOVA shows a significant difference between the means of different segments ($F-statistic=9.95$, $p < 0.05$). The pairwise comparisons reported 2 significant pairs. A statistically significant difference was observed in participants' speed between the pedestrian-shared path and cycle lane ($t-statistic=-4.338$, $p < 0.05$) and between the cycle lane and roadway ($t-statistic=2.80$, $p < 0.05$). However, there was no significant difference in speed between the cycle lane and roadway ($p > 0.05$).

\begin{table}[h]
    \centering
    \captionsetup{justification=centering}
    \caption{Average speed at each route segment. Standard deviation in brackets. \\}
    \label{tab:avg_speed}
    \small
    \begin{tabular}{lccc}
        \toprule
          & \multicolumn{3}{c}{Avg. speed (km/h)}\\
         \cmidrule(lr){2-4} 
         Segment & Mean & Min. & Max. \\
        \midrule
        Pedestrian-shared path & 16.70 (1.84) & 13.21 & 19.84 \\
        Cycle lane & 18.14 (1.00) & 15.82 & 19.65 \\
        Roadway & 17.11 (1.75) & 13.93 & 20.00 \\
        \bottomrule
    \end{tabular}
\end{table}

    \subsubsection{Speed Change Point Detection}
\begin{figure}[h]
  \centering
  \includegraphics[width=0.6\textwidth]{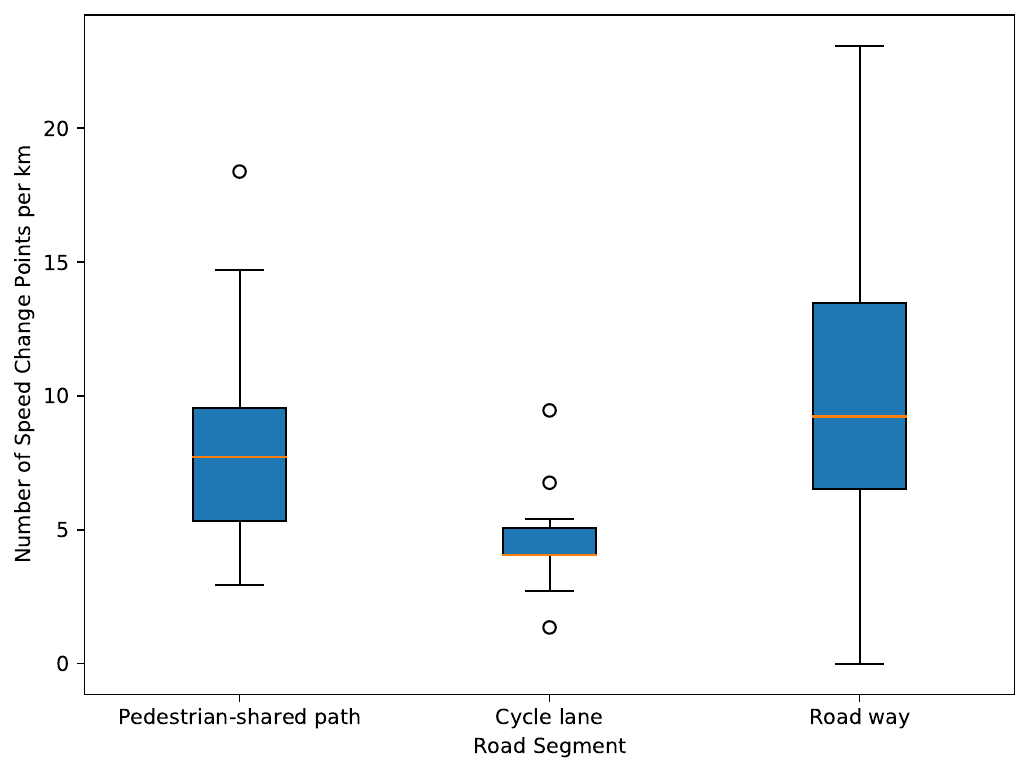}
  \caption{Distribution of Speed Change Points by Segment.}
  \label{fig:speed_change_points}
\end{figure}

To analyse the pattern of speed changes within each route segment, we calculated the number of detected speed change points. Across all participants, 371 speed change points were identified, with the following averages per participant: pedestrian-shared path ($mean = 7.9$, $std. = 4.1$), cycle lane ($mean = 4.3$, $std. = 1.7$), and roadway ($mean = 10.2$, $std. = 5.9$). These change points were triggered by a variety of factors, including encounters with other road users and rider behaviours such as stopping at traffic signals, and navigational turns. Figure \ref{fig:speed_change_points} illustrates the distribution of speed change points across the segments.

The roadway shows the highest median and the highest variation in speed change points, suggesting that riders encounter the most frequent and varied speed adjustments on this type of road. In contrast, the cycle lane reflects a more stable ride, with the lowest variability and the lowest median count of speed change points, indicating fewer speed adjustments. While the pedestrian-shared path shows a moderate mean speed change point value, the variation is less.

        \subsubsection{Gaze Behaviour}
We included gaze data from 14 participants, each with more than 75\% valid gaze samples for our analysis. Six participants were excluded due to insufficient gaze sample rates (below 75\%), and three others were removed because recording sessions were aborted during the study. The issue of data loss due to eye-tracking recording system malfunction is reported in previous studies as well \cite{VONSTULPNAGEL2020222, 10.1145/3204493.3214307}. We annotated data according to 3 segments of the route and calculated fixation count and fixation duration from the unique fixations extracted. To compare the monitoring activities of the participants, we calculated the Euclidean distance from average values for the gaze standard deviation of the participants based on fixations in x and y. Further, the Euclidean distance of standard deviation of the head movements was calculated using accelerometer data (in x, y, and z) extracted from the Tobii eye-tracker. Table \ref{tab:gaze_disp} report the results.

\begin{figure}[h]
    \centering
    \begin{subfigure}{0.5\textwidth}
        \centering
        \includegraphics[width=1\linewidth]{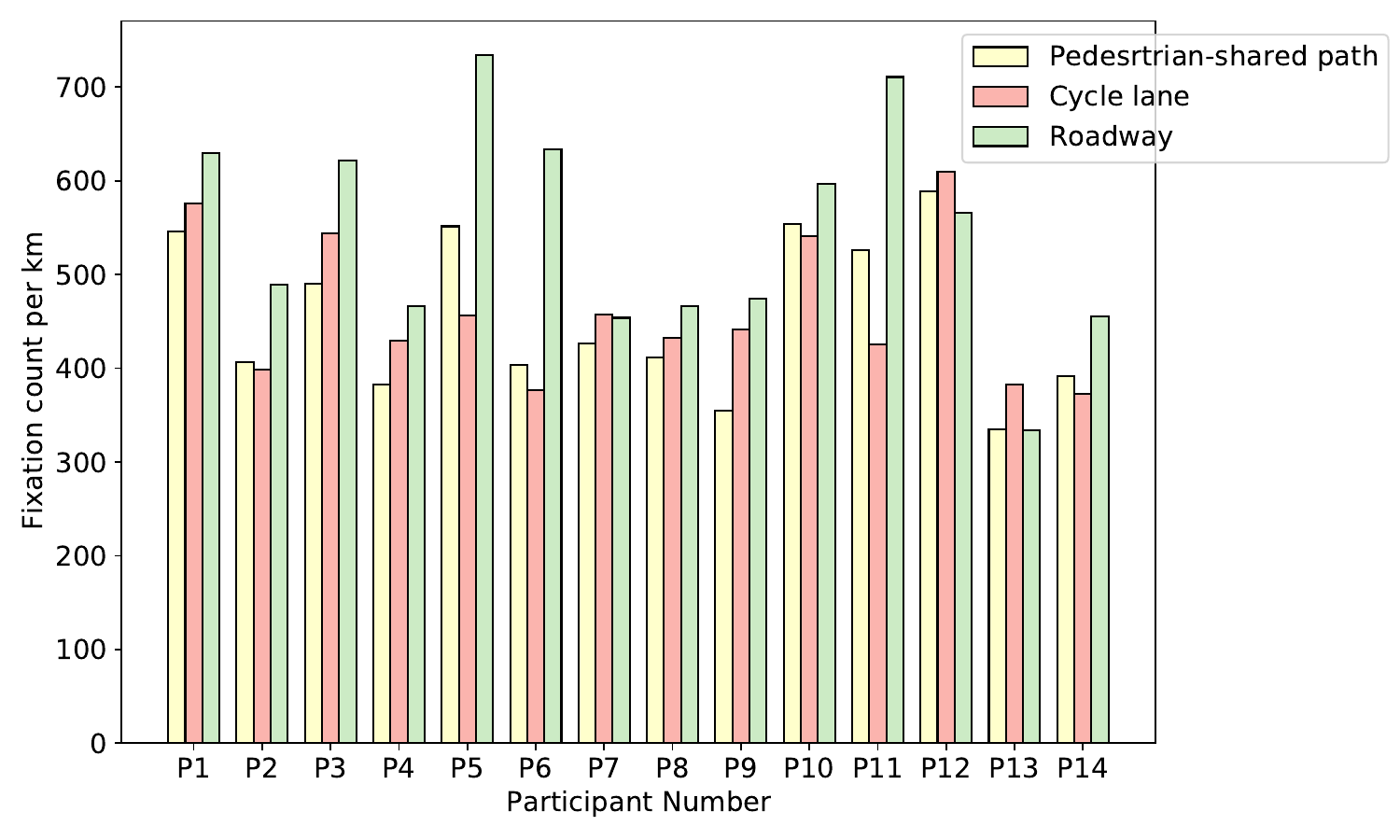}
        \caption{Fixation count distribution}
        \label{fig:Fixation_Counter_distribution}
    \end{subfigure}%
    \hfill
    \begin{subfigure}{0.5\textwidth}
        \centering
        \includegraphics[width=1\linewidth]{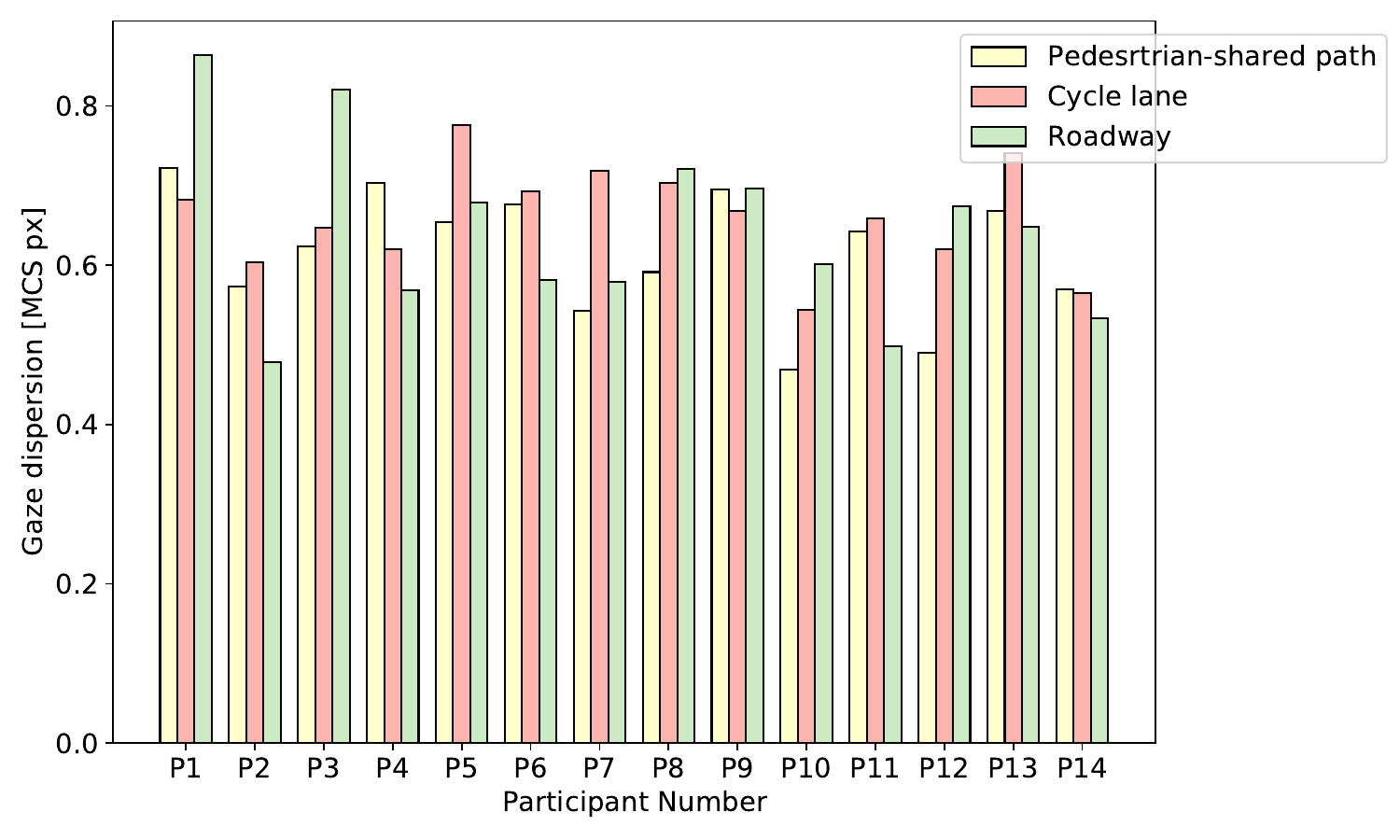}
        \caption{Gaze dispersion distribution}
        \label{fig:gaze_dispersion_distribution}
    \end{subfigure}
    \caption{Distribution of gaze behaviour measures among participants.}
\end{figure}

According to Shapiro-Wilk test, the average fixation count per km was normally distributed. We performed ANOVA with the individual route segments as within-subject conditions. It reported a significant difference between the different route segments ($F-statistic =10.5, p < 0.05$). Pairwise comparison using Tukey's Honest Significant Difference (HSD) test shows a statistically significant higher fixation count in roadway compared to pedestrian-shared path ($ p < 0.05$). Figure \ref{fig:Fixation_Counter_distribution} visualizes the distribution of fixation counts across route segments among participants.

The average duration of fixations was also normally distributed. ANOVA reported a significant difference between route segments ($F-statistic =23.3$, $p < 0.05$), but no specific pairs reached statistical significance in Tukey's HSD.  

According to Shapiro-Wilk test, both gaze dispersion and accelerometer dispersion (head movement) data were normally distributed. ANOVA reported there is no statistically significant difference between the routes. Figure \ref{fig:gaze_dispersion_distribution} shows the distribution of gaze dispersion among participants.

Furthermore, we investigated whether there are variations in monitoring behaviour between frequent and infrequent e-scooter riders. According to the T-test, we found no significant differences in the overall fixation duration between frequent and infrequent e-scooter users ($T-statistic=0.88, p > 0.05$). Similarly, the differences in fixation counts between the two groups -- frequent and infrequent -- were not statistically significant ($T-statistic=0.78, p > 0.05$). The results for gaze dispersion also showed no significant differences between the two groups ($T-statistic=0.22, p > 0.05$). These findings are consistent with those reported in the study by \citet{10.1145/3568444.3568451}, which found no significant differences in the monitoring behaviours of frequent and rare cyclists.

\begin{table}[h]
    \centering
    \captionsetup{justification=centering}
    \caption{Gaze behaviour measures at each route segment. Standard deviation in brackets. \\}
    \label{tab:gaze_disp}
    \small
    \begin{tabular}{lcccc}
        \toprule
          & \multicolumn{4}{c}{Measurement}\\
         \cmidrule(lr){2-5} 
         & Fixations  & Fixation duration  & Gaze dispersion  & Acc. dispersion  \\
        Segment & [count per km] & [ms] & [MCS px] & [m/s2] \\
        \midrule
        Pedestrian-shared path & 454.83 (84.58) & 371.8 (89.3) & 0.61 (0.08) & 0.88 (0.04) \\
        Cycle lane & 460.20 (77.01) & 332.4 (90.4) & 0.66 (0.06) & 0.83 (0.11) \\
        Roadway & 545.05 (113.9) & 295.3 (68.8) & 0.64 (0.11) & 0.86 (0.06) \\
        \bottomrule
    \end{tabular}
\end{table}

{\begin{table}[h]
    \centering
    \captionsetup{justification=centering}
    \caption{Average encounters in each route segment. Standard deviation in brackets. \\}
    \label{tab:avgEncounter_fixation}
     \small
    \begin{tabular}{lrrrr}
        \toprule
          & \multicolumn{3}{c}{Number of encounters per km}\\
         \cmidrule(lr){2-4} 
         Segment &  Pedestrian & Cyclist & Motor Vehicle \\
        \midrule
        Pedestrian-shared path & 31.1 (14.5) & 4.5 (2.8) & 71.8 (20.6)  \\
        Cycle lane & 4.2 (2.9) & 2.4 (2.0) & 83.7 (49.3)  \\
        Roadway & 17.1 (8.4) & 1.4 (1.7) & 187.2 (41.4)  \\
        \bottomrule
    \end{tabular}
\end{table}}

        \subsubsection{Encounters}
\begin{figure}[h]
  \centering
  \includegraphics[width=0.8\textwidth]{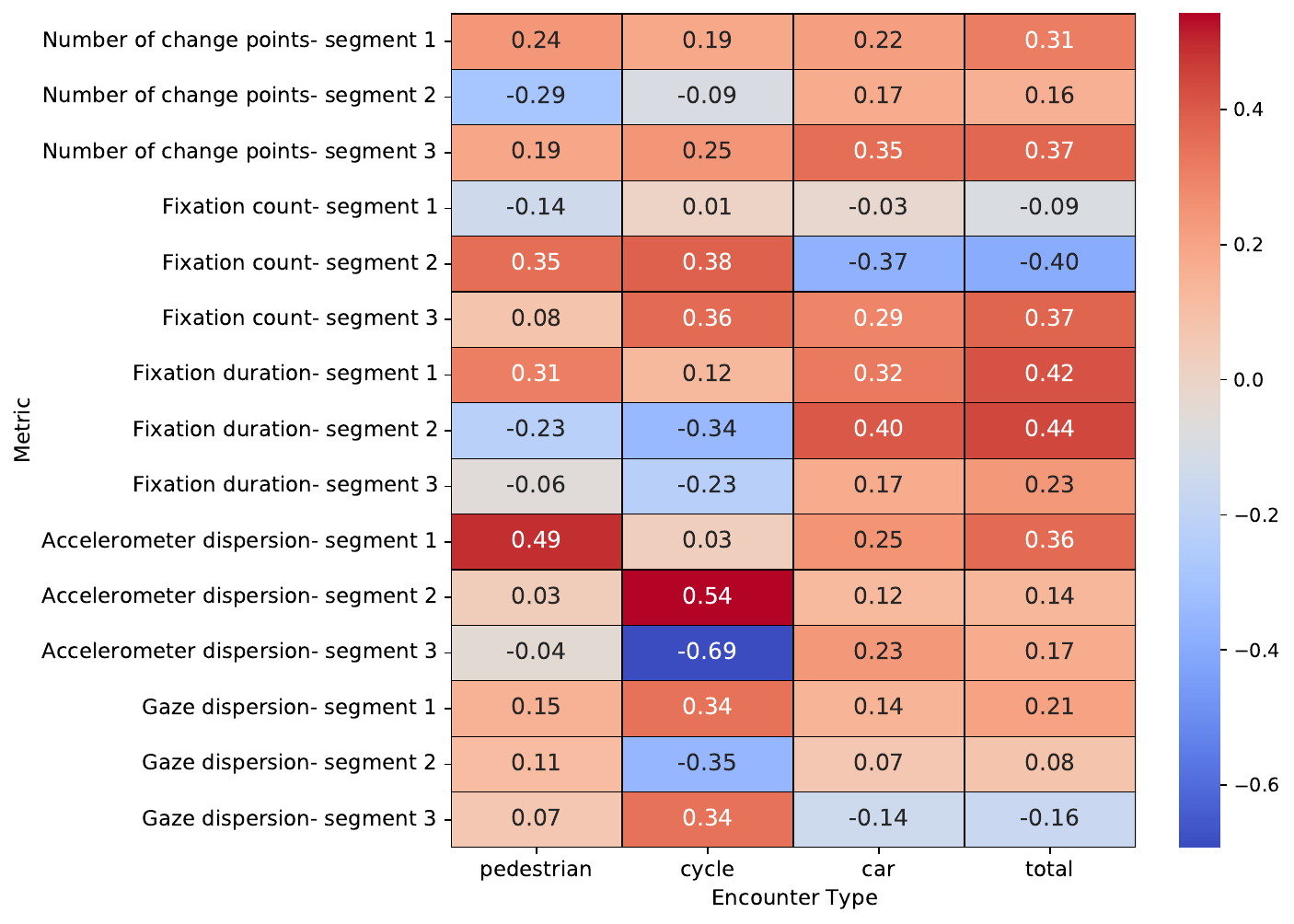}
  \caption{Heat map of correlation values between encounters and metrics. Segment 1: Pedestrian-shared path, Segment 2: Cycle lane, Segment 3: Roadway.}
  \label{fig:correlation_heat_map}
\end{figure}

Using automated video analysis, we calculated the number of pedestrians, cyclists, and cars encountered by participants. The average frequency of encounters in each road segment is presented in Table \ref{tab:avgEncounter_fixation}. Additionally, we analysed the correlation between \textit{encounters} and \textit{gaze behaviour measures} as well as \textit{encounters} and \textit{number of speed change points}. Figure \ref{fig:correlation_heat_map} shows the results across different route segments. In pedestrian-shared path, accelerometer dispersion shows a strong positive correlation with pedestrian encounters. Additionally, speed change points show moderate positive correlations with total encounter types on the pedestrian-shared path. Further, a strong positive correlation between fixation duration and all encounters is visible. 

On the cycle lane, there is a strong positive correlation between accelerometer dispersion and cycle encounters. While the total encounters have a strong positive correlation with fixation duration, cycle encounters and pedestrians encounters are negative. A moderate positive correlations of fixation count with pedestrians and cycles are reported. A moderate negative correlation is shown between fixation counts and car encounters.

On the roadway, a strong negative correlation between accelerometer dispersion and cycle encounters is observed. A moderate negative correlation with cycle encounters and a moderate positive correlation with car encounters can be seen between fixation duration. A moderate positive correlation between overall encounters and fixation counts is seen. Also, on the roadway, moderate positive correlations between speed change points and total encounters are seen.


        \subsubsection{Fixations on AOIs}
\begin{figure}[h]
  \centering
  \includegraphics[width=0.98\textwidth]{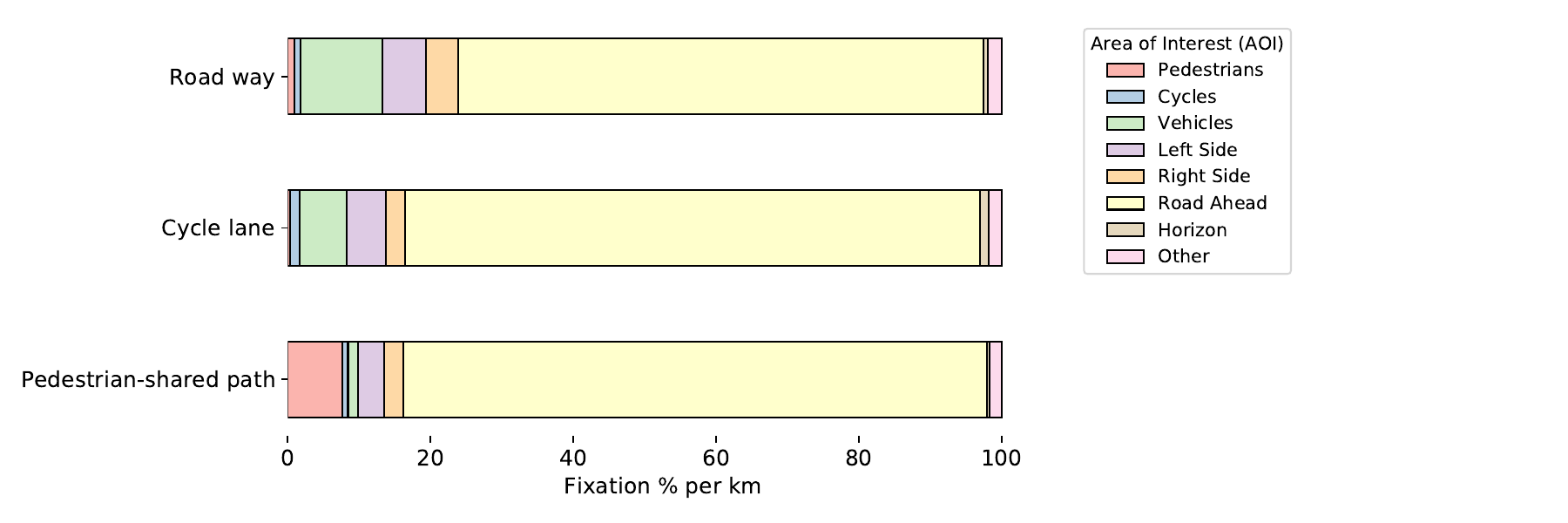}
  \caption{Distribution of fixation percentages in each AOI.}
  \label{fig:Fixation_distribution}
\end{figure}

Figure \ref{fig:Fixation_distribution} shows the distribution of fixations across eight AOIs. Consistent with prior research on gaze behaviour \cite{visualAttenstionTobii, 10.1145/3204493.3214307}, our findings emphasise that the majority of riders' fixations are on the road ahead ($Mean = 78.6\% $). Pedestrian-shared path shows the most varied distribution of fixations, while roadway shows that riders directed more attention on the sides.

\begin{table}[h]
    \centering
    \captionsetup{justification=centering}
    \caption{Average fixation counts at each route segment. Standard deviation in brackets.}
    \label{tab:avg_fixation_count}
    \small
    \begin{tabular}{lrrrrrrrrrr}
        \toprule
          & \multicolumn{8}{c}{Avg. fixation count per km}\\
         \cmidrule(lr){2-9} 
         Segment &  Pedestrian & Cyclist & Motor & Left & Right & Road  & Horizon & Other \\
        &  &  & vehicle & side & side & ahead  &  &  \\
        \midrule
        Pedestrian-shared & 33.2 (12.0) & 3.7 (5.0) & 6.3 (5.1) & 15.8 (12.1) & 11.6 (8.1) & 357.2 (84.6) & 1.9 (2.8) & 7.4 (7.9) \\
        Cycle lane & 1.7 (3.4) & 5.9 (9.5) & 29.3 (21.3) & 24.8 (18.4) & 12.3 (10.3) & 360.8 (66.6) & 5.8 (7.2) & 8.1 (10.9) \\
        Roadway & 4.9 (4.4) & 4.8 (10.5) & 60.3 (26.2) & 32.5 (17.0) & 23.6 (15.5) & 388.9 (107.2) & 3.0 (5.5) & 10.5 (11.0) \\
        \bottomrule
    \end{tabular}
\end{table}

Table \ref{tab:avg_fixation_count} presents the average number of fixations per km for each segment. 
The fixation counts on \textit{Pedestrians}, \textit{Cyclists}, and \textit{Motor vehicles} are affected by the encountered road users. Therefore, excluding them, we performed ANOVA to compare the significant differences between the numbers of fixations at other AOIs. The number of participant fixations on \textit{left side} yielded no significant difference ($F-statistic= 0.45$, $p > 0.05$) across route segments. However, the number of participant fixations on \textit{right side} reported a significant result ($F-statistic= 3.93$, $p =0.031$). For the number of participant fixations on \textit{road ahead}, we report a significant result ($F-statistic= 96.86$, $p < 0.05$). According to Tukey's HSD tests, Fixations on \textit{road ahead} was significantly lower at the pedestrian shared path, than in the roadway and cycle lane. There is no significant difference of participant fixation count on \textit{horizon}, found between the different route segments ($Statistic=1.36$, $p > 0.05$).

    \subsection{Qualitative}
In this section, we present qualitative findings focused on representative encounters between e-scooter riders and other road users. This approach allows us to explore the nuanced interactions and behavioural strategies that emerge in different infrastructure types. Representative encounters were selected based on their relevance to common right-of-way negotiation scenarios, where behaviours such as requesting and negotiating space are frequently observed. This method highlights recurring patterns and unique strategies that may not be fully captured through quantitative data alone, offering a richer understanding of the dynamics involved. 

       \subsubsection{Riders Requesting Right-of-Way} 
\begin{figure}[h]
    \centering
    \begin{subfigure}{0.49\textwidth}
        \centering
        \includegraphics[width=.7\linewidth]{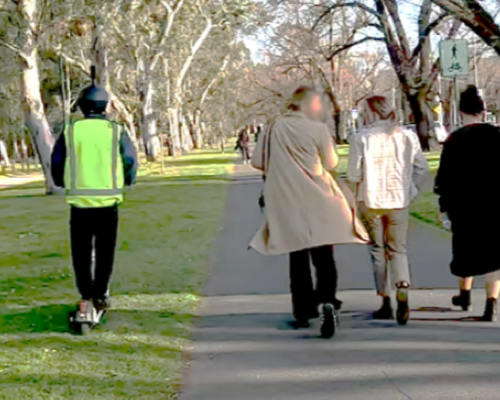}
        \caption{P17 manoeuvre off the road.}
        \label{fig:go_off_road_2}
    \end{subfigure}%
    \hfill
    \begin{subfigure}{0.49\textwidth}
        \centering
        \includegraphics[width=.7\linewidth]{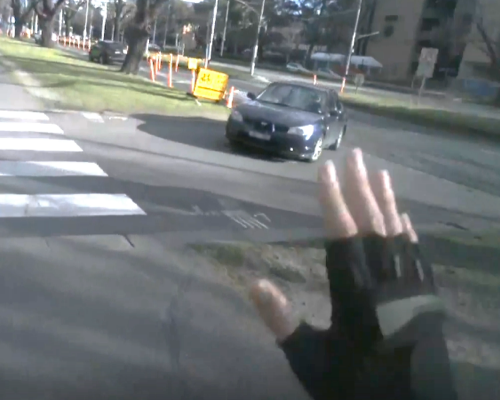}
        \caption{P3 hand signalling car to halt.}
        \label{fig:hand_movement}
    \end{subfigure}
    \caption{E-scooter riders requesting the right-of-way.}
\end{figure}

Notable interactions occurred when groups of pedestrians obstructed the path, leading participants to use various ways of requesting space. A common approach involved ringing the bell to signal their presence. Other riders chose to manoeuvre off the road to avoid direct interference. Figure \ref{fig:go_off_road_2} illustrates P17 moving off-road in response to pedestrians obstructing the way. In addition to these actions, some riders used verbal communication as a form of interaction in shared spaces. For example, P7 expressed gratitude by saying \textit{``thank you''} to two pedestrians who had moved aside, while P2 alerted a group of three pedestrians blocking the path by calling out, \textit{``watch out guys''}. Another noteworthy behavior included the use of hand gestures by e-scooter riders during encounters. As depicted in Figure \ref{fig:hand_movement}, P3 raised his right hand to signal an oncoming motor vehicle to halt and yield the right-of-way. Some riders also used hand signals to indicate turning intentions. Additionally, a few participants opted not to employ any specific strategy, instead waiting patiently and moving slowly until the path cleared. Encounters with children and pets were observed to introduce additional complexities due to the unpredictable nature of their movements.

    \subsubsection{Other Road Users Negotiating Right of Way} 
On the pedestrian-shared path, pedestrians were generally observed to yield to e-scooters upon receiving a request. Many moved to one side to allow the rider to pass or stepped onto the grass in response to the sound of a bell. Some pedestrians guided their companions to the side of the road to clear the path. We also observed some pedestrians used hand gestures to signal the e-scooter rider to proceed.

In the cycle lane, most cyclists overtook the e-scooter riders by using the adjacent lane designated for opposite-direction traffic, often signalling their presence with a bell. Notably, no instances were observed of e-scooter riders overtaking cyclists on this route segment, possibly due to the speed restrictions imposed on shared e-scooters. 

\begin{figure}[htb]
    \centering
    \begin{subfigure}{0.49\textwidth}
        \centering
        \includegraphics[width=0.7\textwidth]{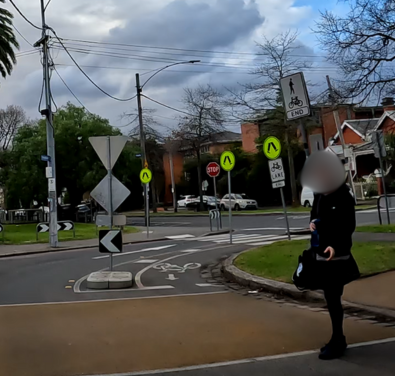}
  \caption{A pedestrian signalled for P23 to proceed.}
  \label{fig:ped-signal-to-go2.png}
    \end{subfigure}%
    \hfill
    \begin{subfigure}{0.49\textwidth}
        \centering
  \includegraphics[width=0.7\textwidth]{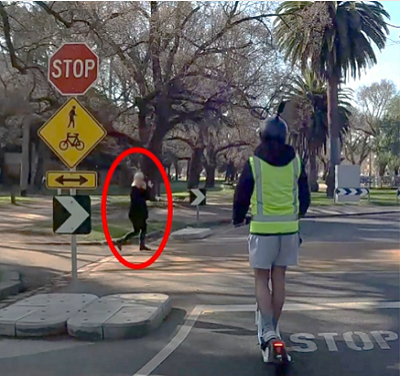}
  \caption{A pedestrian gestured with their hand for P15 to halt.}
  \label{fig:ped_hand_gesture}
    \end{subfigure}
    \caption{Other road users negotiating the way.}
\end{figure}

At a pedestrian crossing on the roadway, interactions varied based on pedestrian intent and perceived readiness to cross. Some pedestrians, upon noticing an approaching e-scooter, allowed the rider to pass (see Figure \ref{fig:ped-signal-to-go2.png}), while others waited for the rider to stop before crossing. For example, as P15 approached a pedestrian preparing to cross, the pedestrian hesitated, ultimately waiting until the rider had significantly reduced speed before stepping onto the road. Once on the road, the pedestrian gestured with a hand signal for the e-scooter rider to halt (see Figure \ref{fig:ped_hand_gesture}), even though the rider had already nearly stopped. The rider then fully stopped, placing one foot on the ground. Eye-tracking data indicated that the rider maintained their gaze on the pedestrian until they safely reached the opposite side to resume his trip.

\section{DISCUSSION}

In this study, we examined e-scooter rider behaviour across three types of mobility infrastructure, identifying differences in monitoring activities, riding speed, and navigational patterns. This section discusses the unique challenges and behaviours exhibited by e-scooter riders and explains how our findings can inform improvements in urban infrastructure design to enhance road safety.

    \subsection{Unique Challenges and Behaviours of E-scooter Riders}

Our results on gaze dispersion reveal no significant differences across route types. However, the highest gaze dispersion was observed on the cycle lane, suggesting that riders are visually scanning a broader area in this environment. Correlation analysis indicates a negative relationship between cycle encounters and gaze dispersion in cycle lanes. This may imply that riders focus more narrowly ahead when other cyclists are present, likely to maintain lane position and avoid collisions. On the roadway, riders exhibit moderate gaze dispersion, suggesting they frequently monitor for immediate hazards within a moderate visual field. This gaze pattern indicates riders focus more intensively on specific regions where immediate action may be necessary. Correlation results show a positive relationship (0.34) with cyclist encounters and a negative correlation (-0.14) with car encounters. This suggests that riders broaden their gaze when encountering other cyclists to stay aware of relative positioning, prioritizing direct observation on the roadway. The narrowest gaze dispersion is reported on pedestrian-shared paths, likely reflecting the less hazardous nature of these routes. Additionally, correlation results indicate that encountering cyclists on pedestrian-shared paths causes riders to broaden their gaze compared to encounters with pedestrians. Overall, gaze dispersion across all three route segments indicates that e-scooter riders pay special attention to cyclists. This attention could be due to the higher speeds of cyclists compared to e-scooters. In the city where this study was conducted, the average cycling speed in urban cycling lanes is between 40 km/h and 50 km/h \cite{Bikelaw}. In contrast, the maximum speed limit for shared e-scooters used in this study is capped at 20 km/h, a restriction enforced in this city and in several other locations where e-scooter trials are conducted \cite{cicchino2024scooter, escooterSpeed}. This speed limit poses a unique challenge for e-scooter riders. Riders have reported feeling vulnerable to aggressive behaviour from motor vehicle drivers, who may not understand that e-scooters are unable to accelerate beyond the enforced speed limit \cite{escooterSpeed}.

In contrast to the findings of \cite{10.1145/3568444.3568451}, which evaluated cyclists’ gaze behaviours, our results did not find a significant difference in accelerometer dispersion across road types. However, the highest accelerometer dispersion was observed on the pedestrian-shared path, reflecting riders’ attempts to maintain awareness of nearby pedestrians. A strong positive correlation (0.49) between pedestrian encounters and accelerometer dispersion on pedestrian-shared paths indicates riders’ need for more frequent head movements to monitor pedestrians in close proximity. The cycle lane showed the lowest accelerometer dispersion, as its predictable, separated environment encouraged riders to maintain a consistent forward gaze with fewer head adjustments. We noted a strong positive correlation (0.54) with cyclist encounters in this lane, reflecting the need to maintain spatial awareness and adjust positioning when sharing the lane with other cyclists. The roadway reported moderate accelerometer dispersion, but correlation results showed a strong negative correlation (-0.69) with cycle encounters, along with a lower correlation between total encounters and head movements. This suggests that riders’ head movements are influenced by factors other than encounters. One potential reason is that riders turn their heads to check for oncoming traffic from behind before merging onto the roadway. This presents a distinct challenge for e-scooter riders, as most shared e-scooters lack turning indicator lights, making it difficult to signal their turning intentions effectively. Although some e-scooter riders use hand signals, this can be risky; releasing the handlebars on an e-scooter may lead to a loss of control \cite{locken2020impact}. In contrast, cyclists commonly use hand signals for turns, which they can perform easily due to the different handling and balance requirements \cite{matviienko2020reminding, hou2020autonomous}. This highlights a key difference in riding practices between e-scooter riders and cyclists.

Further, our observations indicated that interactions between e-scooter riders and pedestrians at crossings often involve a degree of uncertainty. This can be due to the novelty of e-scooters for both riders and other road users, as well as concerns around perceived safety \cite{doi:10.1080/17450101.2021.1967097}. Previous studies \cite{useche2022unsafety, distefano2024comparison} suggest that e-scooter riders are generally perceived as ‘worse’ riders than cyclists. Additionally, findings from \cite{pils2021scooter} reveal a higher acceptance of cyclists over e-scooter riders; for example, car drivers are more likely to yield at cross walks for cyclists than for e-scooter riders. These results underscore the challenges of integrating e-scooters into mixed-use spaces, where they are not yet fully accepted. Another contributing factor to this limited acceptance may be the swerving behaviour often seen in e-scooter riders. In our recordings, we observed e-scooter riders frequently weaving between pedestrians and moving on and off the road. An observational study on bike and e-scooter interactions \cite{distefano2024comparison} highlighted that e-scooters differ fundamentally from bicycles, in mixed traffic, as they are more prone to closer interactions and higher collision risks. Although our study did not specifically examine gaze patterns across varying road conditions, research by \citet{trefzger2021analysis} found that e-scooter riders exhibit more significant shifts in attention on poor-quality roads compared to cyclists, likely due to stronger vibrations and the lack of suspension in e-scooters. 

    \subsection{Infrastructure Development and Route Planning}

Traditional transport infrastructure is typically organized based on mode, for instance, footpaths for pedestrians, cycle lanes for cyclists, and roadways for motor vehicles. The introduction of e-scooters, which currently lack designated lanes, requires them to share spaces traditionally assigned to other modes. Our comparative analysis of e-scooter rider behaviour across various infrastructures, sheds light on determining suitable routes for their use.

Our results indicate that the highest frequency of speed change points occurred on the roadway. Correlation analysis shows a positive correlation (0.37) between speed change point frequency and overall encounters, suggesting that the complexity of the roadway -- with moving vehicles, intersections, and traffic control features such as stop signs and a roundabout -- required riders to frequently adjust their speed. A moderate number of speed change points in the pedestrian-shared lane can be attributed to interactions with pedestrians and cyclists, as supported by correlation results indicating a moderately positive correlation with pedestrians (0.24) and cyclists (0.19). The lowest frequency of speed change points was observed in the cycle lane, where riders maintained a more consistent speed, likely due to fewer interactions and fewer complex obstacles. Additionally, reflecting findings from both cycling \cite{CLARRY2019101594} and e-scooter \cite{cicchino2024scooter} studies, our results showed the highest average speed on the cycle lane. Overall, based on speed analysis, the cycle lane appears to be the most suitable infrastructure for e-scooters.

The number of fixations reveals differences in riders' visual attention across the three infrastructure types. The significantly higher fixation count on roadways suggests that riders continuously assess potential hazards, given their vulnerability to motor traffic. Positive correlations (0.37) with overall encounters indicate that in complex roadway environments, riders must engage in frequent visual checks. The cycle lane shows a moderate fixation count, with correlation results suggesting that encounters with cyclists and pedestrians may have influenced fixations in this setting. On the pedestrian-shared path, a lower fixation count suggests that riders perceive it as a more predictable and safer path. Supporting this finding, \cite{MANTUANO2017408} demonstrated that path segments shared with pedestrians had lower fixation counts compared to designated cycle tracks.

Moreover, the pedestrian-shared path shows the longest fixation duration, suggesting that riders focus on fewer objects and sustain their gaze longer. Correlation results indicate that total encounters (0.42) have a strong positive correlation with fixation duration on this path. In contrast, on the cycle lane, riders engage in shorter, more targeted fixations than on pedestrian-shared paths. According to correlation data, encounters with pedestrians (-0.23) and cyclists (-0.34) have negative correlations with fixation duration, while total encounters show a strong positive correlation (0.44). Minimal pedestrian encounters and the strategy of quick scanning for other cyclists could explain this. The shortest fixations are reported on the roadway, where the dynamic environment -- with moving and parked cars, intersections, and traffic control features -- likely leads riders to scan rapidly across various AOIs. Our findings align with previous research showing longer fixation durations on paths shared with pedestrians compared to separate cycle paths \cite{MANTUANO2017408}. Additional studies provide evidence of riders focusing on parked cars \cite{10.1145/3607822.3614532, venkatachalapathy2022naturalistic}.

Analysis of AOIs reveals that fixations on the road ahead are significantly higher across all three route segments, likely as a precautionary measure to anticipate potential hazards. This is consistent with prior findings on cyclists \cite{10.1145/3204493.3214307}. Further, we found that \textit{road ahead} fixations are significantly lower on pedestrian-shared path compared to cycle lane and roadway. This suggests that, in the presence of pedestrians, riders allocate more visual attention to their surroundings, reducing their focus on the road directly ahead. Similarly, \citet{MANTUANO2017408} observed that cyclists’ balance of visual attention shifts when pedestrians are present. This pattern indicates that e-scooter riders may recognize pedestrians as vulnerable road users and adjust their gaze to monitor them more closely.

Overall, our findings demonstrate the benefits of dedicated cycle lanes, separated from both motor vehicles and pedestrians, as suitable spaces for e-scooter use. Previous studies have shown dedicated bicycle lanes attracted more e-scooter traffic over roadway \cite{YANG2022204, ZHANG2021102761, ghaffar2023meta, WEISS2024100047} and are generally perceived as safer by riders \cite{ZHANG2021102761, LAA2020102874}. Additionally, research indicates that e-scooter riders prefer to be separated from pedestrians \cite{hardinghaus2022attractive, WEISS2024100047}. Given the similar riding behaviours of e-scooter riders and cyclists -- particularly regarding speed and spatial awareness -- integrating e-scooters into existing cycling infrastructure could be an effective solution. These findings suggest that urban planners should prioritize the implementation of dedicated cycle lanes within city layouts. Furthermore, to enhance safety on mixed-use paths, strategies like clear signage, appropriate regulations, and public awareness campaigns are essential. Advances in technological solutions, such as rider assistance \cite{riderAssistance}, collision detection \cite{2024_Yan}, and footpath detection systems \cite{footpathDetection1,footpathDetection2} could also play a crucial role in fostering safer interactions among all road users.

\section{LIMITATIONS AND FUTURE WORK}

While we took steps toward understanding micro-mobility user behaviour across various types of transport infrastructure in real-world conditions, we acknowledge several limitations. Our sample included only 23 participants, with additional reductions in usable data for speed (19 participants) and eye-tracking (14 participants) analyses due to device malfunctions. Moreover, for participant safety, the selected route was limited to park surroundings, which may not fully reflect the complexities of e-scooter interactions in busier urban or crowded settings. Future studies should investigate riding behaviour in a broader, more diverse participant group and in higher-density urban environments.

Furthermore, the precaution of participants wearing high-visibility jackets were necessary, the attire could have altered the typical behaviours of the other road users, making them more aware and potentially more cautious than they would be otherwise. However, in the city where we conducted the study, it is not uncommon for cyclists to wear high-visibility jackets for added safety.

Finally, our methodology could be strengthened by incorporating a post-study phase that combines real-time video analysis with immediate post-ride interviews. This approach would allow for a deeper exploration of participant perspectives, offering insights into the rationale behind specific riding choices made during the study.

\section{CONCLUSION}
In this paper, we investigated e-scooter riding behaviours across different types of transport infrastructure: pedestrian-shared paths, cycle lanes, and roadways. Our naturalistic study involved 23 participants navigating a predefined route while equipped with various ubiquitous computing devices (e.g., cameras, a bike computer, eye-tracking glasses). Employing a combination of established data analysis techniques, we gained novel insights into e-scooter rider behaviour. Our findings on speed variation, gaze patterns, and navigational practices reveal unique challenges faced by e-scooter riders, including difficulty keeping up with faster-moving traffic (particularly cyclists and motor vehicles) due to the capped speed limit on shared e-scooters, challenges in safely indicating turns due to balance risks with hand signals, and limited acceptance from other road users in mixed-use spaces. Additionally, our study highlights the benefits of dedicated cycle lanes -- separated from motor vehicles and pedestrians -- as suitable spaces for e-scooter use.

\begin{acks}
This research was conducted by the ARC Centre of Excellence for Automated Decision-Making and Society (project number CE200100005), and funded by the Australian Government through the Australian Research Council. We gratefully acknowledge Lime Network Pty Ltd for providing e-scooters for our study.
\end{acks}

\bibliographystyle{ACM-Reference-Format}
\bibliography{sample-manuscript}


\end{document}